\documentclass[final,1p,times]{elsarticle}

\usepackage{amssymb}

\usepackage{ulem}
\usepackage{lineno}
\usepackage{verbatim}

\journal{Physics Letters B}

\begin{document}

\begin{frontmatter}

\title{Evidence for Shape Co-existence at medium spin in $^{76}$Rb}

\author[Yor]{R.~Wadsworth\corref{cor1}}
\author[Lun]{I.~Ragnarsson}
\author[Lun]{B.G.~Carlsson}
\author[Lun]{Hai-Liang~Ma \fnref{Bei}}
\author[Yor]{P.J.~Davies}
\author[Gue]{C.~Andreoiu \fnref{SFU}}
\author[Ham]{R.A.E.~Austin \fnref{Hal}}
\author[Arg]{M.P.~Carpenter}
\author[Ray]{D.~Dashdorj}
\author[Arg]{S.\ J.\ Freeman \fnref{Man}} 
\author[Gue,Van]{P.E.~Garrett}
\author[Arg]{J.~Greene}
\author[Ced]{A.~G\"{o}rgen}
\author[Yor]{D.G.~Jenkins}
\author[Yor]{F.~Johnston-Theasby }
\author[Yor]{P.~Joshi}
\author[Ber]{A.O.Macchiavelli}
\author[Arg]{F. Moore}
\author[Arg]{G.~Mukherjee \fnref{Cedfn}}
\thanks{Present address: Variable Energy Cyclotron Centre, Kolkata 700064, India}
\author[Was]{W. Reviol}
\author[Was]{D.G.~Sarantites}
\author[Arg]{D.~Seweryniak}
\author[Gue]{C.E.~Svensson}
\author[Gue]{J.J.~Valiente-Dob\'{o}n \fnref{Len}}

\address[Yor]{Department of Physics, University of York, Heslington,
York YO10~5DD, UK}
\address[Lun]{Department of Mathematical Physics, Lund Institute of Technology, P.O. Box 118 S-221 00, Lund, Sweden}
\address[Gue]{Department of Physics, University of Guelph, Guelph,
Ontario N1G 2W1, Canada}
\address[Ham]{Department of Physics and Astronomy, McMaster University, Hamilton, Ontario L8S 4K1, Canada}
\address[Arg]{Physics Division, Argonne National Laboratory, 9700 South Cass Avenue,
Argonne, IL 60439}
\address[Ray]{North Carolina State University, Rayleigh, North Carolina, 27695}
\address[Van]{TRIUMF, 4004 Wesbrook Mall, Vancouver, BC, V6T 2A3, Canada}
\address[Ced]{CE Saclay, Daphnia/SphN, 91191 Gif-sur-Yvette Cedex France}
\address[Ber]{Lawrence Berkeley National Laboratory, Berkeley, CA 94720}
\address[Was]{Department of Chemistry, Washington University, St. Louis, MO 63130}

\cortext[cor1]{Principle corresponding author}

\fntext[Bei]{Present address: Dept. of Nuclear Physics, China Inst. of Atomic Energy,
  P.O. Box 275(18), Beijing 102413, China}
\fntext[SFU]{Present address: Dept. of Chemistry, Simon Fraser University, Burnaby, Bristish Columbia,
Canada V6T 1Z4}
\fntext[Hal]{Present address: Saint Mary's University, Halifax NS B3H 3C3, Canada}
\fntext[Man]{Present address: Department of Physics and Astronomy, 
University of Manchester, Manchester, M15 9PL, UK}
\fntext[Len]{Present address: INFN, Laboratory Nationali di Legnaro, I-35020, Italy}
\fntext[Cedfn]{Present address: CE Saclay, Daphnia/SphN, 91191 Gif-sur-Yvette Cedex France}

\date{\today}

\begin{abstract}

Four previously known rotational bands in $^{76}$Rb have been extended to moderate spins 
 using the Gammasphere and Microball $\gamma$ ray
and charged particle detector arrays and the $^{40}$Ca($^{40}$Ca,3pn) reaction
at a beam energy of 165 MeV.  
 The properties of two of the negative-parity bands can only readily be interpreted in terms
of the highly successful Cranked Nilsson-Strutinsky model calculations if they have
the same configuration in terms of the number of g$_{9/2}$ particles, but
they result from different nuclear shapes (one near-oblate and the other
near-prolate). 
These data 
appear to constitute a unique example of shape co-existing structures at
medium spins.

\end{abstract}


\end{frontmatter}

\newpage


One of the remarkable properties of the nuclear quantal many body system is its ability to
minimize its energy by adopting different nuclear shapes for a relatively small
cost in energy compared to the total binding energy. This, coupled with the shape driving effects
of the odd nucleons can, in certain circumstances, result in different nuclear shapes
being possible at low excitation energies in nuclei.
Recent calculations \cite{möl09} have pinned down
the regions of the nuclear chart where nuclei may assume different shapes
at close to groundstate energies. In some cases as many as four different
minima, corresponding to different shapes, are found in the potential energy
surface for a single nucleus. The degree to which the groundstate wavefunction becomes 
a mixture of these differently shaped states or if the mimima give rise to 
individual nuclear states is still an open and interesting question.

 Various reviews have been performed over the last 20 years on shape co-existence in atomic
nuclei, with the most recent being by Heyde and Wood \cite{hey10}. Potentially, one of best examples of
nuclear shape coexistence at low spins is found in the light lead nucleus, $^{186}$Pb, where
the first three excited 0$^{+}$ states are believed to result from spherical,
oblate and prolate minima, respectively, in the potential energy surface \cite{and00}. 
This particular phenomenon of shape coexistence has been discussed in terms of intruder states based
on proton particle-hole excitations across the Z=82 shell gap \cite{hey10,woo92,woo99}. 
The phenomenon is also known to occur at much higher spins where 
many excited rotational bands can often be found. Since each band is usually built on different 
excited configurations this often results in different shapes that can change with spin. 
It is however only when the states are sufficiently 
similar and possess the same quantum numbers that one may see an
interaction between rotational bands. Such interactions can then provide information about the degree of
shape mixing.

The neutron deficient nuclei with mass
$A=70-80$ are located in a region of large deformed shell gaps
\cite{naz85}.
Strong gaps exist at both oblate
(34, 36) and prolate (36,38) nucleon numbers \cite{Bou03}. The first evidence for shape co-existence at low spins in
this region was proposed in $^{72}$Se \cite{ham74} and a detailed review of all the early data
was made by Wood et al. \cite{woo92} some years later.
More recently detailed studies of shape co-existence and mixing between the low-spin oblate and prolate states
has been investigated in $^{72-78}$Kr \cite{Bou03}.
$^{76}$Rb, which has 37 protons and 39 neutrons is located in the
region of interest. However, whilst $^{76}$Rb does not directly possess any of the nucleon numbers where 
large shell gaps occur for oblate shapes, there are long standing suggestions for the presence
of shape coexistence in the nucleus at low spins \cite{hof85} as well as indications for
evidence of the phenomenon in N=39 isotopes of Ge, Se, Kr and Sr (see fig 38 of ref
\cite{hey10}).
At moderate spins the nucleus has been found to be dominated by 
five very regular rotational structures \cite{har95}.
In the present work these structures have been extended to spins of the order of 30$\hbar$, with
over 40 new $\gamma$ rays being observed. The structures observed, and the interactions between them
can be interpreted, with the aid of cranked Nilsson-Strutinsky (CNS) calculations, as 
providing evidence of an excellent example of the coexistence at moderate spins of oblate and prolate structures.
Of particular interest is the fact that two bands can be interpreted as being constructed from the 
same basic configurations, i.e. with the same number of particles excited to the
g$_{9/2}$ shell, but with very different shapes. 
To our knowledge, this is the first example of this kind of shape coexistence
at high spin where the rotational bands in both minima are observed in an
extended spin range. A further interesting feature is that it has been
possible to extract approximate interaction matrix elements between the
bands in the two minima at spin values $I \approx 12$ and $I \approx 20$,
respectively. 

\vspace{0.5cm}

The experiment was performed at the Argonne National Laboratory using the ATLAS accelerator to 
produce a $^{40}$Ca beam at 165 MeV. This beam was used to bombard a 350 $\mu$g/cm$^2$ 
$^{40}$Ca target
 that was flashed on both sides with 150$\mu$g/cm$^2$ of gold to prevent 
oxidation.  The reaction channel of interest in the present work was  
 $^{40}$Ca($^{40}$Ca,3pn)$^{76}$Rb. $\gamma$ rays from $^{76}$Rb were detected 
using the Gammasphere array \cite{Lee:90}, which consisted of 99 Compton suppressed Ge detectors, 
and identification of evaporated charged particles was performed using the Microball  array \cite{Sar:96}.  An event 
was triggered on the condition of at least four of the HPGe detectors firing in prompt 
coincidence. A total of 1.5x10$^9$ high-fold events were recorded.
The information from the Microball array, that provides the energies and directions of the
detected charged particles, allowed for an offline 
event-by-event reconstruction of the momenta of the residual nuclei \cite{ler03, Sew:94}, thereby 
resulting in an improved energy resolution for the $\gamma$-ray  peaks. 
 
\begin{figure}
\includegraphics[width=1.0\columnwidth]{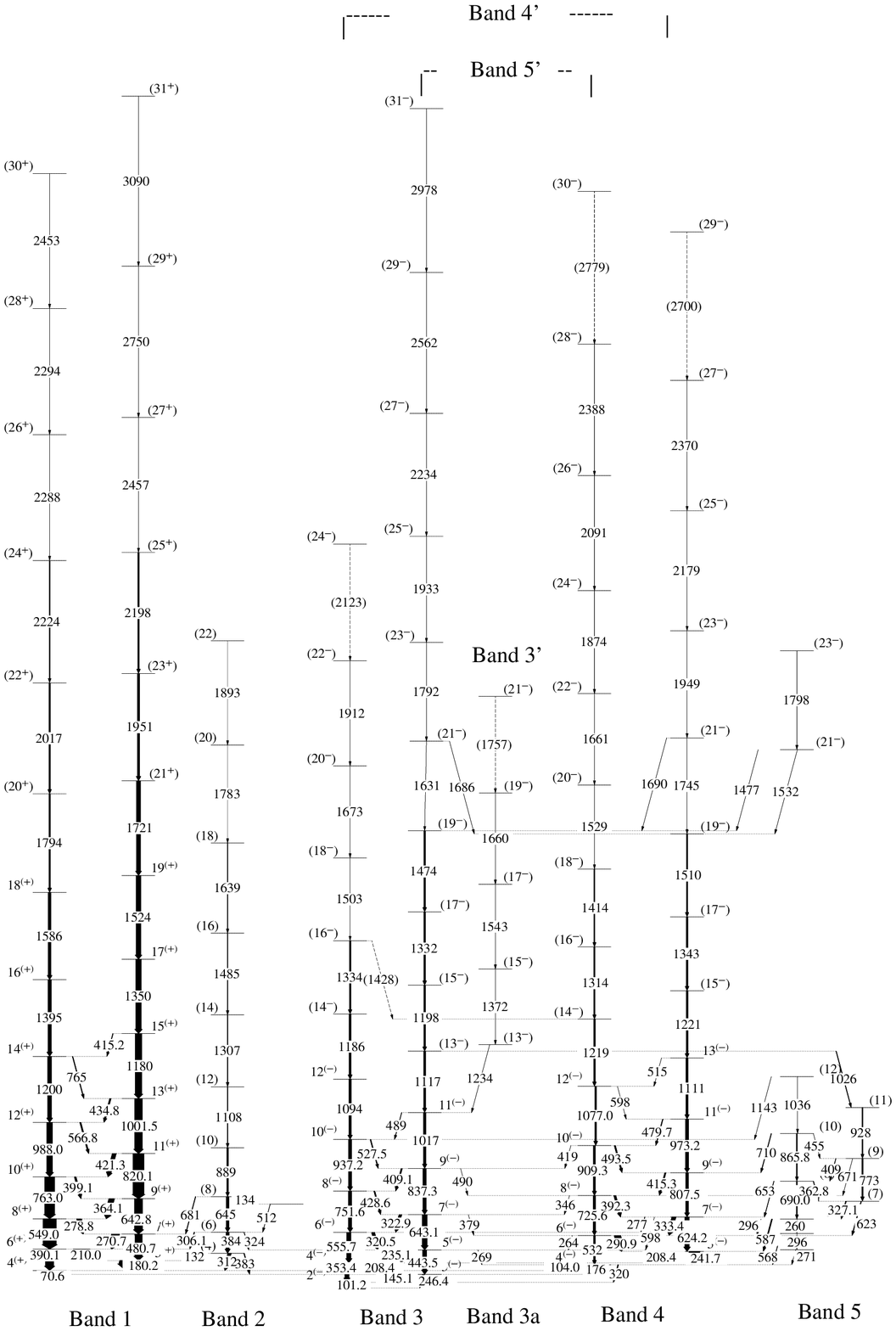}
\caption{Energy level scheme for $^{76}$Rb proposed from a combination of the work by Harder
et al \cite{har95} and the present work. The negative parity bands are
labelled as bands 3, 4, 5 and 3a
according the the strongest B(E2) transitions,
whilst at higher spins it is
indicated how the bands labelled as 3, 4 and 5 at low spin
develop into the high spin bands 3', 4' and 5'
as discussed from the band mixing calculations. The exchange of character
occurs for spin values $I=10-16$ as illustrated
in Fig. \ref{cross} below. 
}
\label{levelscheme}
\end{figure}

Events coinciding with the detection of no $\alpha$ particles and 1 or 2 protons were used to help
select the nucleus $^{76}$Rb. Attempts to enhance the channel of interest (3pn) using the 3p-gated
data and the total energy plane (TEP) 
method \cite{Sve:97} did not improve the signal to background significantly. This is due to the
finite charged particle detection efficiencies and the strongly overlapping locations in the
3p-gated TEP of the 3pn events in which the neutron is not detected, 4p events in which 
one of the protons was not detected, and $\alpha$3p events in which the $\alpha$-particle was not detected.
An open TEP gate was thus utilised in order to maximise statistics.  
The events, with the above particle gating conditions,
 were used to create a  $\gamma$-$\gamma$-$\gamma$ cube which
was used to extend the known \cite{har95} energy level decay scheme for the nucleus. The
level scheme, deduced with the aid of the RADWARE \cite{rad95} 
graphical analysis package, is shown in Fig. \ref{levelscheme}. The spins and parities of the extended rotational
structures are assigned on the assumption that the observed transitions are stretched E2's, since
it was not possible to obtain sufficiently accurate directional correlation from oriented state (DCO)
values to unambiguously confirm these.

\begin{figure}

\includegraphics[angle=-90,width=1.0\textwidth]{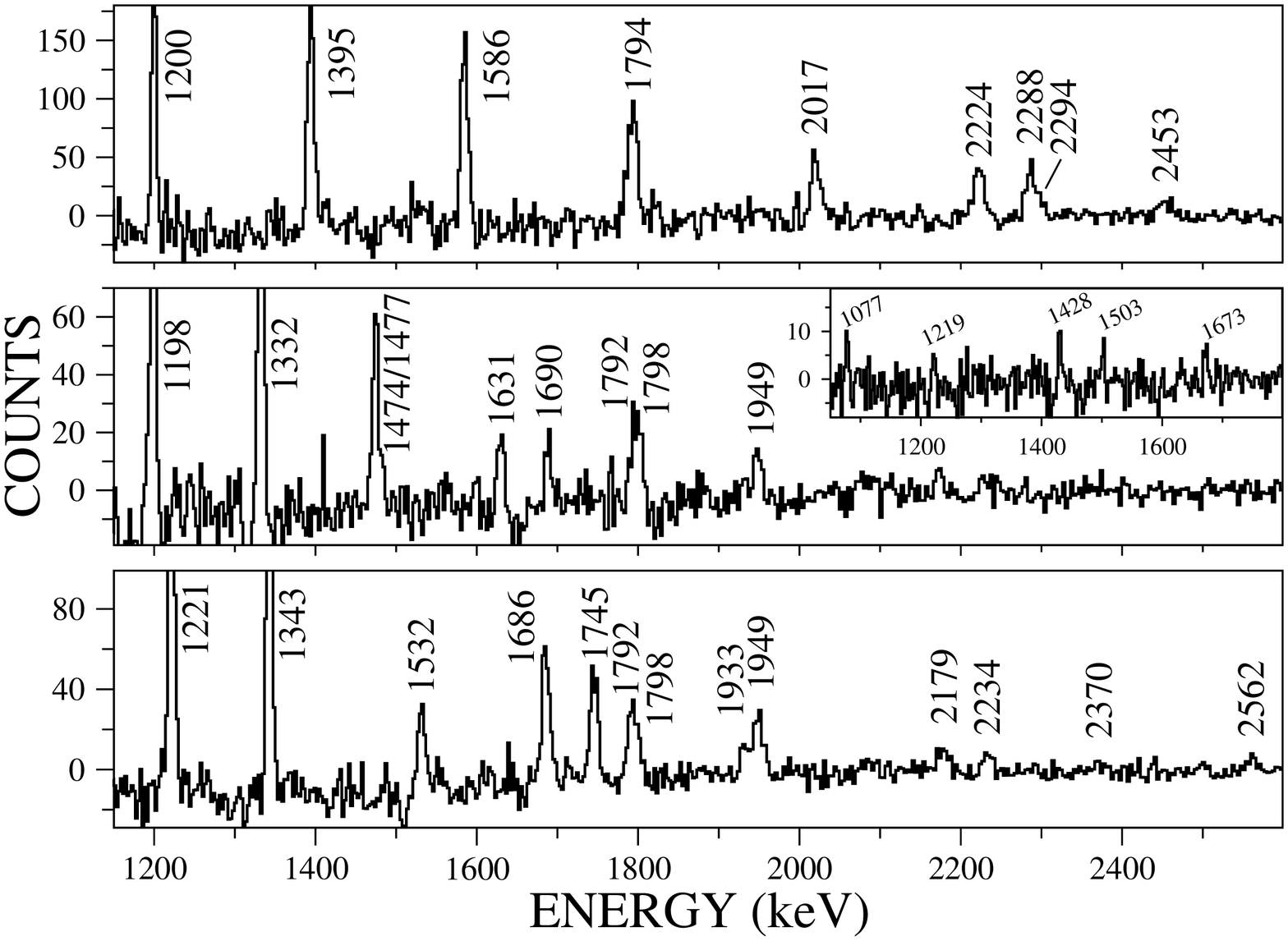}
\caption{Coincidence spectra obtained from the E$_{\gamma}$-E$_{\gamma}$-E$_{\gamma}$
cube discussed in the text. The top spectrum represents a sum of double gates between all transitions in 
the even spin sequence of band 1 from 1200 keV up to 2453 keV. The middle spectrum shows a sum
of double gates between all transitions from the 246 keV $\gamma$ ray up to the 1332 keV
$\gamma$ ray and the 1474/1477 keV transitions in the odd spin sequence of band 3. The inset to
this spectrum was created from a sum of double gates between the 909 and 1077 keV transitions
of the even spin sequence of band 4 and the 1503 and 1673 keV $\gamma$ rays of the even spin sequence 
of band 3. The bottom spectrum was created from a sum of double gates between all transitions from the
242 to 1343 keV transitions in the odd spin sequence of band 4 against the 1510 keV $\gamma$ ray.}
\label{spectra}
\end{figure}

Fig. \ref{spectra} shows some partial spectra in support of the proposed level scheme and to illustrate 
the overall level of statistics and quality of the data. 
The double gating conditions used to create the spectra shown
from the E$_{\gamma}$-E$_{\gamma}$-E$_{\gamma}$ cube are described in the relevant figure captions.
The present work agrees with the previous work \cite{har95}
for the low spin part of the decay scheme but most bands have
been extended to considerably higher spin values. 
The $\alpha$ = 0, 1 signatures of `band 1' with positive parity
have been extended by 6 and 5 transitions, respectively.
The negative parity bands are labelled as bands 3, 4 and 5 at low-spin.
In the present work the
even and odd spin sequencies of band 3 have been extended by 5 and 8 transitions, respectively.
A new sequence of 4 (plus 1 tentative) transitions with $\alpha$ = 1 (band 3a'), has also 
been observed in the present work
which feed into the original $\alpha$ = 1 structure of band 3 at spin 11$^{(-)}$. It is
assumed that all the observed transitions in band 3a' have E2 multipolarity. It is also
interesting to note that band 3 is connected to band 5, the relevance of which will be
discussed further below. Finally, band 4 has been extended by 4 (plus 1 tentative) 
and 5 transitions for the $\alpha$ = 0, 1 signatures, respectively. 

A closer study of the band structure in Fig. \ref{levelscheme} indicates
an interesting sequence of interactions between the negative parity structures. 
It is evident that in the $I=20-30$ spin range, the odd
spin sequence of band 3 is signature degenerate with the even spin sequence
of band 4 (cf. Fig. \ref{expth} below), i.e. these two sequences must be signature partners which is in
contradiction to the band assignment at low spin 
according to Ref.\ \cite{har95} and as drawn in Fig. \ref{levelscheme}. Furthermore, 
the odd spin sequences of bands 3 and 4 interact around $I=20$ with connecting
transitions which are similar in strength to the in-band transitions. A possible
idea to solve the problem would be to interchange these two
sequences for spin values $I \geq 21$. However, the present 
way of connecting the bands leads to a much more smooth behaviour as a function of spin than
would be the case if the two sequences were interchanged. We also note
that there is a weak transition connecting the even spin sequences of these two
bands, i.e. a transition from the $I=16$ state of band 3 to the $I=14$ state
of band 4 (this connection was first identified after
the theoretical analysis suggested that a strong interaction should exist between the
two structures). Thus, a possible (and perhaps more logical)
alternative would be to interchange the even spin sequences
of these two bands  for spin values $I \geq 14$.

\begin{figure}
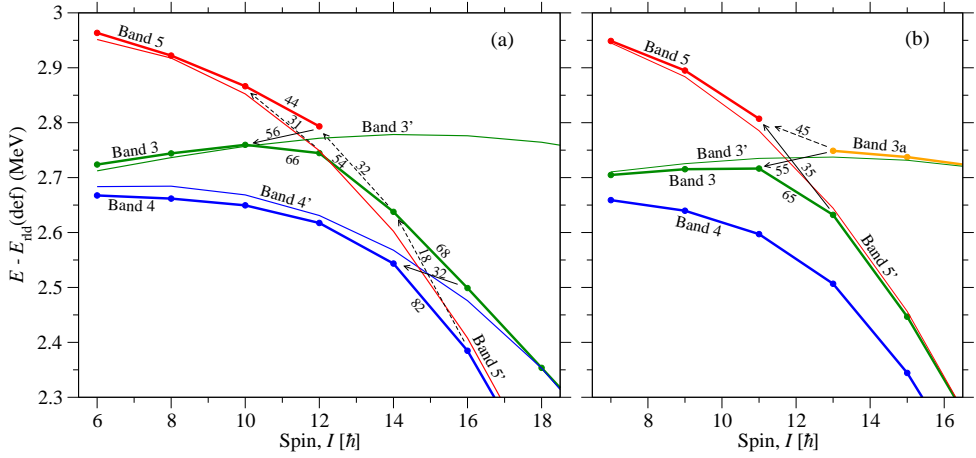

\includegraphics[clip=true,height=6cm]{cross1-mod.eps}
\includegraphics[clip=true,height=6cm]{cross3-mod.eps}
\caption{Illustration of the three-band crossing for the even spin 
(left panel) and the two-band crossing for the odd spin
(right panel) negative
parity states in $^{76}$Rb, with the energies drawn relative to
a rotating liquid drop reference. The thick lines 
with filled circles show the bands as they
are drawn in Fig. \ref{levelscheme}. The thin lines define the 
smooth unperturbed bands which results in a very good description of the observed
states when they interact with constant matrix elements. 
B(E2) branching ratios
calculated assuming constant B(E2)'s within the smooth bands and no
connecting B(E2)'s are written out around the crossings,
where observed and unobserved transitions are shown using arrows with 
filled and dashed lines, respectively.}
\label{cross}
\end{figure}

The crossings between the even spin negative parity states are
illustrated in Fig. \ref{cross}(a). A closer look
at the observed states in that figure does not only suggest the crossing
between bands 3 and 4 around $I=14$ but also a crossing between bands 3 and
5 around $I=11$. In order to find out if this scenario appears
consistent with the observed bands,  we have carried out a three-band-mixing 
calculation. In this calculation, three smooth 
unperturbed bands, labelled as 3', 4' and
5' are parameterized with a
moment of inertia with a linear dependence on $I$.
They interact with strengths
which are assumed to be constant over the spin range
considered, $I=6-20$.
With a least-square fit, all observed states in
this spin range are reproduced within $\pm 5$ keV
with an interaction matrix element of 16 keV for the 3' and 5' bands that cross 
at $I \approx 11$
and 44 keV for the 4' and 5' bands which cross at $I \approx 15$. The 
interaction matrix element 
between the 3' and 4' bands comes out as 28 keV but it appears less well determined.
Branching ratios were also calculated assuming the same transition strengths,
B(E2)'s, occur within the smooth unperturbed
bands (this assumption is roughly consistent with our interpretation
below, however, at this stage it seems reasonable not to assume any specific
assignment)
and no transition probabilities connecting these bands .
With these assumptions the calculated branching ratios do show reasonable agreement with experiment
in that the observed bands follow the strongest calculated branching
ratios, see Fig. \ref{cross}(a). For the crossing between bands 3 and 4 
(or 5' and 4') in Fig. \ref{cross}a at
$I \approx 15$, the strongest connecting transition
with a predicted 32\% branching ratio is observed whilst the one with a
predicted 18\% branch is not observed. At the $I \approx 11$ crossing,
the $I=12$ state of band 5 is predicted to decay to two  $I=10$ states with
about equal probability (44\% and 56\%) and both these transitions are
observed, while for the decay of the  $12^+$ state of band 3, it is only the
transition that is predicted to be strongest (66\%) which is observed. 

Band mixing calculations were also performed for the odd spin sequences.
However, in this case, the two crossings between bands 3 and 5 and bands 3 and 4 are
found to be too displaced in spin ($I \approx 12$ and $I \approx 20$, respectively)
to make it possible to fit smooth bands with our simple formula in the spin
range covering both band crossing regions. We therefore carried out simple two-band mixing calculations at
each of the respective crossings. For the crossing at $I \approx 12$, band 3a is also
involved, i.e. two bands can be formed if the $I=1-11$ spin range of band 3 is combined
with band 3a ($I=13-21$) and band 5 ($I=7-11$) is combined with the 
$I=13$ and upwards spin range of band 3, see Fig. \ref{cross}(b).
With the non-interacting 
bands parameterized as discussed above and with a coupling strength of 38 keV it is
again possible to fit all states in the $I=7-17$ spin range within $\pm 5$
keV. Furthermore, the observed bands follow the strongest predicted
transitions where, for example, as indicated in Fig. \ref{cross} 
the $I=13$ state of band 3 
is calculated to have a 65\% branching
within the band but also a 35\% branching to the $I=11$ state of band 5,
which appears consistent with the fact that both transitions are
observed experimentally. Furthermore, 
the $I=13$ state of band 3a is predicted to have its
strongest branch (55\%) to the
$I=11$ state of band 3, in agreement with observation, however, the band mixing calculation
also indicates that there should be a strong branch (45\%) to the $I=11$ state of band 5,  
suggesting that it should be possible to also observe this transition. An extensive search
has not been able to locate evidence for this decay, indicating that this
particular decay is much weaker than predicted.

The experimental states at the crossing between the odd spin sequences in bands 3 and 4 at
$I \approx 20$ (not shown in Fig. \ref{cross})
are fitted reasonably well by two unperturbed bands
parameterized as above.
The best fit is obtained
using an interaction strength of 14 keV but the fit 
shows little improvement compared with the assumption of no
interaction between the bands.
With the 14 keV interaction,
the inband transitions
are predicted to be about 3 times as strong as the connecting transitions.
The observed transitions are so weak that it is difficult to determine their
relative intensities but the prediction appears consistent with the
observed decays from the $21^-$ state of band 4 while the two transitions from
the $21^-$ state of band 3 have roughly the same intensities.

The new level scheme of $^{76}$Rb will now be compared
with the predictions of the CNS model \cite{BR.85,PhysRep,Car06},
which does not include pairing.
The CNS theoretical approach has been used very successfully to describe the high-spin 
rotational structures in other nuclei in this mass region, e.g. see 
\cite{PhysRep,Kel02,AF.05,Val:05a,Val:05,Car06b,Dav10}.
Thus, one may expect that similar calculations for the high spin states in $^{76}$Rb may 
provide a good description of the observed structures in this nucleus.

The lowest energy collective states of $^{76}$Rb which are
calculated in the CNS model are shown in 
the middle panels of Fig. \ref{expth}
relative to a rotating liquid drop reference. 
The collectivity is essentially governed by the number of particles
excited from the $N=3$ orbitals below the $Z=N=40$ gaps 
to the g$_{9/2}$ orbitals above these
gaps. Thus, the bands in Fig. \ref{expth} have at least
7 particles excited. An interesting feature of the calculations is that there are
several less collective configurations with fewer particles 
excited which are predicted to exist in the yrast region. 
However, no such states have been observed in the
present experiment. 
The reason for this may lie in the fact that it is much easier to identify smooth
collective structures that are close to yrast at high-spin where population occurs
in fusion evaporation reactions, but another possibility
is of course that the less collective structures 
are predicted to occur at too low an excitation energy in the calculations.

The bands in the upper panels of Fig. \ref{expth} are the unperturbed
bands 3', 4' and 5', defined at high spin in Fig.  \ref{levelscheme} and
in the interaction region ($I \approx 6-16$) in Fig. \ref{cross}. They connect
smoothly to the observed bands in the low spin region, where the primed bands 
coincide with the unprimed bands. 
These bands are those which appear to correspond to
the evolution of smooth structures from low to high spin and they are thus the ones that
naturally correspond to the CNS configurations.
 
As exemplified for the [3,4] 
configuration in Fig. \ref{pes},
most of the calculated collective configurations show  coexistence between
close-to-oblate and close-to-prolate shape with a well-defined barrier
between the two minima. (The [p,n] notation represents the number of g$_\frac{9}{2}$
protons and neutrons in the configuration.) Consequently, we expect well defined individual bands
in the two minima.
In order to find interpretations for
the observed bands,
we have followed the lowest positive parity and
three lowest negative parity bands of each signature
in the middle panel of Fig. \ref{expth}. 
Considering first the positive parity band, band 1, this is well
described by  the close-to-prolate [3,5] configuration which is calculated to be lowest
in energy. A particularly interesting feature is the discontinuity 
(band crossing) that is observed for the $\alpha$ = 0
signature in the experimental energy shown in the upper panel.
Noting that in the cranking model, we do not try to describe 
the exact energies in the crossing region, it is very satisfying 
that the observed discontinuity  ties in very well with a predicted
discontinuity in the the $\alpha$ = 0 signature of the [3,5] configuration,
see middle panel of Fig. \ref{expth}. 
The discontinuity
in the calculated band arises from the crossing between the negative signature of the
third g$_{9/2}$ neutron orbital and the lowest
d$_{5/2}$g$_{7/2}$ orbital. 

For negative parity, the two lowest observed bands, 4' and 5' at high-spin, are shown
in the upper right panel of Fig. \ref{expth}. 
These bands are very well described by the
two lowest calculated bands which correspond to close-to-oblate 
and close-to-prolate
shape of the [3,4] configuration, see middle panel and compare
with Fig. \ref{pes}. Note especially that 
in both experiment and calculations, the lowest band (band
5' assigned to the prolate shape) is almost 
signature degenerate while the $\alpha = 1$ signature
is clearly favoured in
the next lowest band (band 4' assigned to the oblate shape).
The third lowest band of negative parity, band 3', 
is shown in the left panels of Fig. \ref{expth}. For this band,
it
is only the $\alpha = 1$ branch which is observed up to
$I \approx 20$ so we cannot draw any decisive conclusions but it appears to be 
reasonably well described
by the third lowest calculated band of negative
parity, which is the near-prolate band
with a  [4,5] configuration.

With the assignments specified above, the differences between
calculated and experimental energies of the states are shown in the lower panels of
Fig. \ref{expth}. For spin
values $I \gtrsim 15$, these differences are
considerably smaller than the expected \cite{Car06}
accuracy of $\pm 1$ MeV.
Furthermore, the differences are similar for all the bands,
i.e. they come close to 0.5 MeV for $I \approx 15$ and lie roughly
in the range  0-0.5 MeV for $I=25-30$. This latter fact
indicates that the relative properties of the bands are 
well reproduced in the calculations.

\begin{figure}
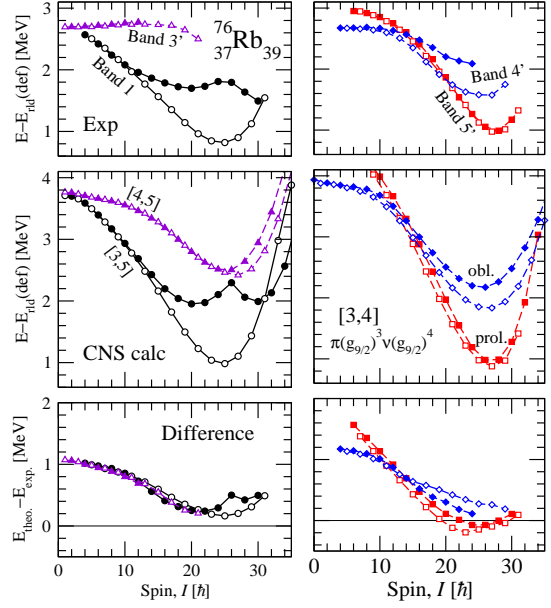

\begin{center}
\includegraphics[clip=true,height=8cm]{expthar-mix.eps}
\includegraphics[clip=true,height=8cm]{expthbr-mix.eps}
\caption{Observed (upper panels) and calculated (middle panels) rotational bands
in $^{76}$Rb and their difference (lower panels), with the two lowest negative
parity bands which show shape coexistence in the panel to the right and the
lowest positive parity band and third lowest negative parity band to the
left. The observed bands are drawn as the smooth bands 3', 4' and 5' in
the interaction region, see Fig. \ref{cross} 
and connect smoothly to the observed states at
low and high spin, see text for details.
Filled (open) 
symbols are used for even (odd) spin states and full (dashed) lines for
positive (negative) parity.}
\label{expth}
\end{center}
\end{figure}

\begin{figure}
\begin{center}
\includegraphics[clip=true,height=6cm]{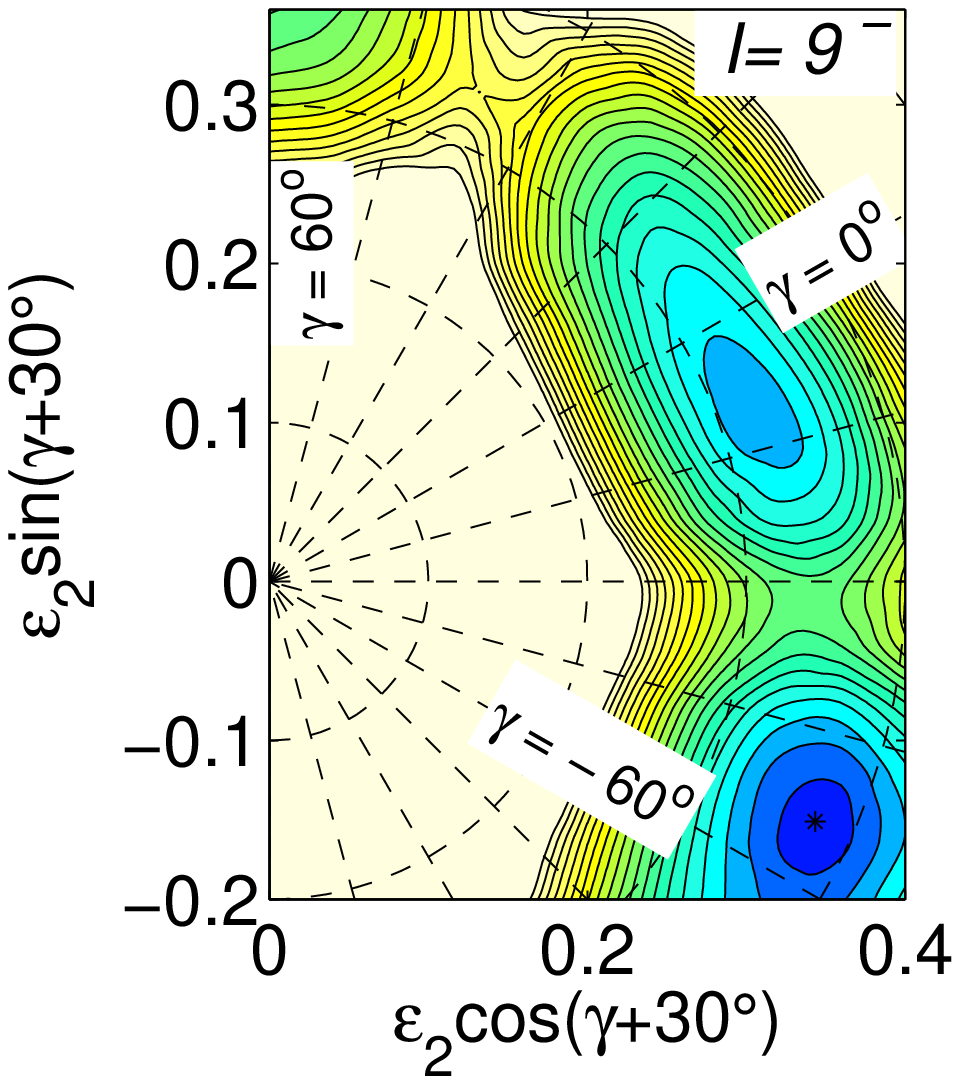}
\includegraphics[clip=true,height=6cm]{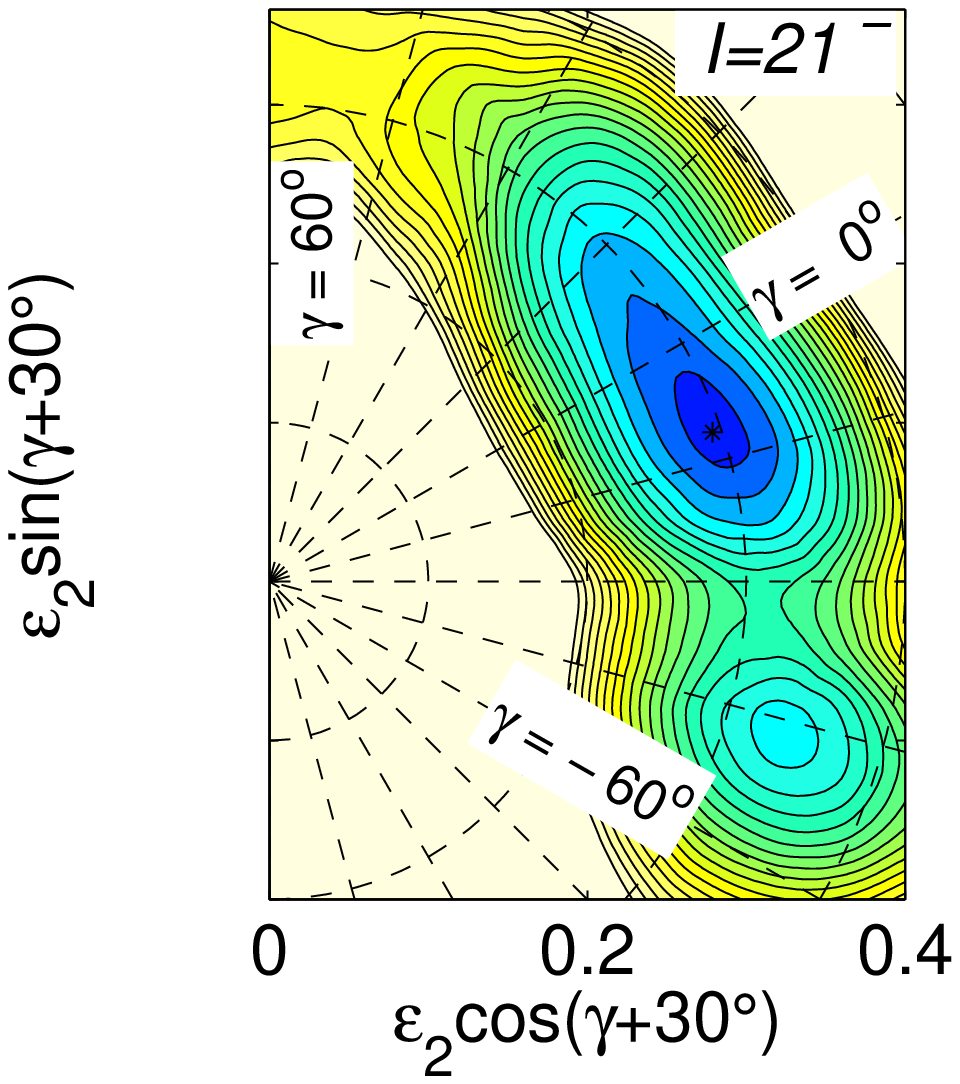}
\caption{Calculated potential energy surfaces for the 
$I = 9$ and $I = 21$ states of the [3,4] configuration. The contour line
separation is 0.2 MeV.}
\label{pes}
\end{center}
\end{figure}

The differences in the lower panels of Fig. \ref{expth} become more
positive for low spin values indicating the importance of the pairing energy, which
is not included in the CNS calculations. These pairing energies are
however small for the odd-odd $^{76}$Rb nucleus, a feature which can be
concluded independent of our interpretation from the fact that
the moments of inertia for the different bands come close
to the rigid body value at low spin values. However, in order to evaluate this further
we have also carried out calculations including pairing
in the formalism presented in Ref. \cite{Car08}, with
particle number projection and with energy minimization not
only in the shape degrees of freedom, $\varepsilon_2$, $\gamma$ 
and $\varepsilon_4$ but also in the pairing degrees of freedom,
$\Delta$ and $\lambda$. In agreement with the discussion above,
the contributions from pairing are found to be small 
at low spin values and they decrease with increasing
spin. These pairing energies will not change the
general structure which means that, for example, the potential
energy surfaces with pairing included 
are found to be very similar to those shown in Fig. \ref{pes}
with two well separated minima. This supports our analysis 
which we have preferred to carry out mainly in
the unpaired formalism, making our conclusions
more transparent, mainly due to the unique possibilities
to fix configurations in an extended spin range. 

The most interesting result in our analysis is the coexistence of 
close-to-prolate and close-to-oblate bands of both signatures
for the [3,4] configuration. This coexistence is illustrated in the 
potential energy surfaces of Fig. \ref{pes} and is seen 
to give an excellent
description of the observed bands. Furthermore, it is only when two
coexistent shapes are considered for the [3,4] configuration that it 
appears possible to find a reasonable explanation of the band referred to as 4' in Fig. 1
with its large signature splitting.
In view of the excellent description 
within the CNS approach of
the level schemes in neighbouring nuclei \cite{AF.05,Val:05a,Car06b,Dav10}
and also in other regions of the
($Z,N$) chart, we consider that this conclusion about coexistent shapes in
the $I=15-30$ spin range is very reliable. Considering neighboring nuclei,
we have also carried out CNS calculations for $^{77}$Rb which results in
a description of all high-spin bands with a similar accuracy to that achieved in $^{76}$Rb,
but where only one band is observed for each configuration. It is also 
interesting to note that the `new type of band
crossing' discussed for $^{77}$Rb in Ref \cite{Har96}, 
is straightforward to understand in the CNS formalism 
as a crossing between bands with a different number 
of high-$j$ particles, see e.g. Ref \cite{Ben85b} where such crossings in the
h$_{11/2}$ shell are discussed. 

For $^{76}$Rb, it is gratifying that
the observed and calculated bands in the $I=10-15$ spin range cross in very
much the same way in experiment and calculations. This is especially true
for bands 4' and  5' which cross around $I=15$ in a very similar way
to the oblate and prolate bands of the [3,4] configuration that has been assigned to
them, see Fig. \ref{expth}. This makes it tempting to also
speculate about the presence of shape coexistence around the band heads, but a more
thorough investigation of such a scenario is outside the scope of the
present letter. The band mixing calculations show that the even spin
states of the oblate and prolate structures interact with a matrix element 
close to 50 keV at $I \approx 15$ while the odd spin states interact with
a smaller matrix element which is less well determined around $I=20$.
This can be compared with the much larger interactions for the 
coexistent shapes at low spin for $^{72-78}$Kr \cite{Bou03}, 
suggesting a decreasing matrix element with increasing spin.
This is also consistent with theoretical calculations 
\cite{Koi11} including
large amplitude vibrations, which indicate that the spread of the
wave-function over different shapes reduces with increasing
angular momentum.   

It is also interesting to consider how the bands evolve
for spins higher than presently observed. At high spin, all
configurations have a tendency to approach termination, i.e. the shape
trajectories will slowly approach the non-collective limit
at $\gamma = (-120, 60^{\circ})$ corresponding to the left border
of the energy surfaces in Fig. \ref{pes}.  
As seen in 
Fig. \ref{expth}, the  close-to-oblate branch of the [3,4] configuration
is predicted to cross the  close-to-prolate branch for spin values just
beyond $I=30$. The reason appears to be larger components of 
(holes in) the f$_{7/2}$ shell which thus have an important contribution
when building
spin for $I \gtrsim 30$. Indeed, with no contribution from the 
f$_{7/2}$ shell, the maximum spin in the $[3,4]$ configuration is
$I_{max} =35$ but no real termination is predicted at this spin
value, e.g. see \cite{Val:05, Dav10}.

\vspace{0.5cm}

In summary, high-spin states have been populated in $^{76}$Rb using the $^{40}$Ca($^{40}$Ca,3pn) reaction.
This has led to extensions of all the previously know rotational bands.
>From the way the three negative parity bands (of both signatures)
interact, they are redefined into structures which evolve smoothly with
spin where approximate interaction strengths between these structures have been
extracted. 
They are compared with Cranked Nilsson-Strutinsky
(CNS) calculations which provide an excellent description of all
bands. In particular, two of the bands can be understood as being built from the same
configuration (in terms of the number of g$_{9/2}$ particles), but arising from different -
near coexistent - close-to-prolate and close-to-oblate nuclear shapes. This appears to
constitute the best example of such coexistent structures of similar
configurations at medium spin. The comparison between the observed and
calculated bands suggests that the coexistence can be followed down to the
band heads where, contrary to the $I=20-30$ spin range, the close-to-oblate
structure is calculated to be lowest in energy.  Finally, considering the special features of
band crossings and coexistence observed in $^{76}$Rb it would be very interesting to
carry out a new experiment to get a more complete understanding of these
features; hopefully also making it possible to extend the bands to higher
spins where their termination or non-termination \cite{Val:05, Dav10} is
another important fingerprint of their configurations.

\section*{Acknowledgements}
\label{Acknowledgements}

This research was partially supported by the U.K. Science and 
Technology Research Council, by the Swedish Research Council, by The
Natural Science and Engineering Research Council 
of  Canada,  and the U.S. Department of Energy under Contract Numbers DE-AC02-06CH11357
and DE-FG02-88ER-40406.

\label{Bibliography}
\bibliographystyle{phaip}

\begin{thebibliography}{99}

\bibitem{möl09} P. M\"{o}ller {\em et al.}, Phys. Rev. Lett. {\bf 103}, 
212501 (2009).

\bibitem{hey10} K. Heyde and J.L. Wood, 
\newblock submitted to Rev. Mod. Phys. (2011).

\bibitem{and00} A.N. Andreyev {\em et al.}, 
\newblock Nature {\bf 405}, 430 (2000).

\bibitem{woo92} J.L. Wood, K. Heyde, W. Nazarewicz, M. Huyse and P. Van Duppen,
\newblock Phys. Rep. {\bf 215}, 101 (1992).

\bibitem{woo99} J.L. Wood, E.F. Zganjar, C. De Coster and K. Heyde, 
\newblock Nucl. Phys. {\bf A651}, 323 (1999).

\bibitem{naz85} W. Nazarewicz, J Dudek, R. Bengtsson, T. Bengtsson and I. Ragnarsson, 
\newblock Nucl. Phys A {\bf 435}, 397 (1985).

\bibitem{Bou03}  E. Bouchez  {\em et al.} Phys. Rev. Lett {\bf 90}, 028502
(2003); E. Clement {\em et al.}, Phys. Rev. C {\bf 75}, 054313, (2007).

\bibitem{ham74} J.H. Hamilton et al., Phys. Rev. Lett. {\bf 32}, 239 (1974).

\bibitem{hof85} S. Hofmann, I. Zychor, F.P. Hessberger and G. Munzenberg, 
\newblock Z. Phys. {\bf A325}, 37 (1986).

\bibitem{har95} A. Harder {\em et al.}, 
\newblock Phys. Rev. C {\bf 51}, 2932 (1995).

\bibitem{Lee:90} I.-Y ~Lee, 
\newblock Nucl. Phys A {\bf 520}, 641c (1990).

\bibitem{Sar:96} D. G.~Sarantites {\em et al.}, 
\newblock Nucl. Instrum. Meth.  {\bf A383}, 506 (1996).

\bibitem{ler03} F. Lerma {\em et al.}, 
\newblock Phys. Rev.  {\bf C67}, 044310 (2003).

\bibitem{Sew:94} D.~Seweryniak {\em et al.}, 
\newblock Nucl. Instr. and Meth.  {\bf A340}, 353 (1994).

\bibitem{Sve:97} C.E~Svensson {\em et al.}, 
\newblock Nucl. Instrum. Meth. A {\bf 396}, 288 (1997).

\bibitem{rad95} D. Radford, 
\newblock Nucl. Instr. Meth.  A {\bf 361}, 297 and Nucl. Instr. Meth. A 
{\bf 361}, 306 (1995).

\bibitem{BR.85} T.\ Bengtsson, and I.\ Ragnarsson, Nucl.\ Phys. A {\bf 436}, 14 (1985).

\bibitem{PhysRep} A.\ V.\ Afanasjev, D.\ B.\ Fossan, G.\ J.\ Lane and I.\ Ragnarsson,
Phys.\ Rep. {\bf 322}, 1 (1999).

\bibitem{Car06} B.\ G.\ Carlsson and I.\ Ragnarsson,
Phys.\ Rev. {\bf C74}, 011302(R) (2006).

\bibitem{Kel02} N.\ S.\ Kelsall {\em et al.}, 
\newblock Phys. Rev.  {\bf C65}, 044331 (2005).

\bibitem{AF.05} A.\ V.\ Afanasjev, and S.\ Frauendorf, 
\newblock Phys. Rev.  {\bf C71}, 064318 (2005).

\bibitem{Val:05a} J J.~Valiente-Dob\'{o}n {\em et al.}, 
\newblock Phys. Rev. {\bf C71}, 034311 (2005).

\bibitem{Val:05} J J.~Valiente-Dob\'{o}n {\em et al.}, 
\newblock Phys. Rev. Lett. {\bf 95}, 232501 (2005).

\bibitem{Car06b}
B.G.\ Carlsson and I.\ Ragnarsson, Proc.\ `Frontiers in Nuclear Structure,
Astrophysics and Reactions, FINUSTAR', Kos, Greece, Sept.\ 12-17, 2005
(eds.\ S.\ Harissopulos, P.\ Demetriou  and 
R.\ Julin) AIP Conf.\ Proc.\ 831 (2006) p.\ 60. 

\bibitem{Dav10} P.J. Davies {\em et al.},  Phys. Rev. C {\bf 82}, 061303, (2010). 

\bibitem{Car08} B.\ G.\ Carlsson {\em et al.},
Phys.\ Rev. {\bf C78}, 034316 (2008).

\bibitem{Har96} A. Harder {\em  et al.}, 
Phys. Lett. B {\bf 374}, 277 (1996).

\bibitem{Ben85b} T.\ Bengtsson and I.\ Ragnarsson,
Phys. Lett. {\bf 163B}, 31 (1985).


\bibitem{Koi11} K. Sato and N. Hinihara, Nucl. Phys. A {\bf 849}, 53 (2011).



\end{thebibliography}

\end{document}